\documentclass[twocolumn,showpacs,preprintnumbers,amsmath,amssymb]{revtex4}
\usepackage{graphicx}
\usepackage{dcolumn}
\usepackage{bm}
\usepackage{subfigure}
\usepackage{color}

\begin{document}

\title{A fresh look at double parton scattering}

\author{M.G.~Ryskin}
\affiliation{Petersburg Nuclear Physics Institute, Gatchina, St. Petersburg, 188300, Russia}

\author{A.M.~Snigirev}
\affiliation{D.V. Skobeltsyn Institute of Nuclear Physics, M.V. Lomonosov Moscow State University, 119991, Moscow, Russia }

\date{\today}
\begin{abstract}
A revised formula for the inclusive cross section for double parton scattering in terms of the modified collinear two-parton distributions extracted from deep inelastic scattering is suggested. The possible phenomenological issues are discussed.
\end{abstract}
\pacs{12.38.-t, 13.85.-t, 13.85.Dz, 14.80.Bn}


\maketitle
\section{\label{sec1}Introduction}
The presence of hard multiple parton interactions in high-energy hadron-hadron collisions has been convincingly  de\-mo\-nstrated by the AFS~\cite{AFS}, UA2~\cite{UA2}, CDF~\cite{cdf4jets,cdf}, and D0~\cite{D0} Collaborations, using events with the four-jets and $\gamma+3$-jets final states, thus, providing new and complementary information on the proton structure.

In general, four high-$E_T$ jets (or $\gamma+3$-jets) may be produced either in the collision of one pair of partons in $2\to 4$ hard subprocess or via the simultaneous interaction of two parton pairs, that is in two $2\to 2$ subprocesses. In the last case in each dijet (or $\gamma+$ jet) system the transverse momenta of two high-$E_T$ jets more or less balance each other.
 
The possibility of observing two separate hard  collisions was proposed long ago. Their theoretical investigation has a long history and goes back to the early days of the parton model~\cite{landshoff,takagi,goebel} with subsequent extension to perturbative QCD~\cite{paver,humpert,odorico,Mekhfi:1983az,Ametller:1985tp,Halzen:1986ue,sjostrand,Mangano:1988sq,Godbole:1989ti,Drees:1996rw,trelani,trelani2,sjostrand2,sjostrand3,strikman,Gaunt:2009re,Calucci:2010wg,Blok:2010ge,Strikman:2010bg,dremin,Ceccopieri:2010kg,del,DelFabbro:2002pw,Hussein:2006xr,Hussein:2007gj,Kulesza:1999zh,Cattaruzza:2005nu,maina,Berger:2009cm,Bahr:2008dy,corke,Gaunt:2010pi,Maina:2010vh,Diehl:2011tt,stir}. 

Nevertheless, the phenomenology of multiple parton interactions relies on the models that are physically intuitive but involve significant simplifying assumptions. The\-re\-fore, it is extremely important to combine theoretical efforts in order to achieve a better description of multiple interactions, in particular, double scattering, which will be the dominant multiple scattering mode at the LHC. In this letter we consider some steps towards this purpose. The cross section formulae currently used to calculate double scattering processes are revised in terms of the modified collinear two-parton distributions extracted from deep inelastic scattering (DIS).

Let us recall, that with only the assumption of factorization of the two hard parton processes $A$ and $B$, the inclusive cross section of a double parton scattering process in a hadron collision is written in the following form
\begin{eqnarray} 
\label{hardAB}
\sigma^D_{(A,B)} = \frac{m}{2} \sum \limits_{i,j,k,l} \int \Gamma_{ij}(x_1, x_2; {\bf b_1},{\bf b_2}; Q^2_1, Q^2_2)\nonumber\\
\times\hat{\sigma}^A_{ik}(x_1, x_1^{'},Q^2_1) 
\hat{\sigma}^B_{jl}(x_2, x_2^{'},Q^2_2)\nonumber\\
\times\Gamma_{kl}(x_1^{'}, x_2^{'}; {\bf b_1} - {\bf b},{\bf b_2} - {\bf b}; Q^2_1, Q^2_2)\nonumber\\
\times dx_1 dx_2 dx_1^{'} dx_2^{'} d^2b_1 d^2b_2 d^2b,
\end{eqnarray}
where ${\bf b}$ is the usual impact parameter, that is the distance between centres of incoming (the beam and the tar\-get) hadrons in trans\-verse plane. $\Gamma_{ij}(x_1, x_2;{\bf b_1},{\bf b_2}; Q^2_1, Q^2_2)$ are the double parton distribution functions, depending on the longitudinal momentum fractions $x_1$ and $x_2$ and on the  transverse position ${\bf b_1}$ and ${\bf b_2}$ of the two parton undergoing the hard processes $A$ and $B$ at the scales $Q_1$ and $Q_2$. $\hat{\sigma}^A_{ik}$ and $\hat{\sigma}^B_{jl}$ are the parton-level subprocess cross sections. The factor $m/2$ is a consequence of the symmetry of the expression for interchanging parton species $i$ and $j$. $m=1$ if $A=B$ and $m=2$ otherwise.

The double parton distribution functions $\Gamma_{ij}(x_1, x_2; {\bf b_1},{\bf b_2}; Q^2_1, Q^2_2)$ are the main reason of interest in multiple parton interactions. These distributions contain in fact all the information of probing the hadron in two different points simultaneously, through the hard processes $A$ and $B$.

It is typically taken that the double parton distribution functions may be decomposed in terms of longitudinal and transverse components as follows:
\begin{eqnarray} 
\label{DxF}
\Gamma_{ij}(x_1, x_2;{\bf b_1},{\bf b_2}; Q^2_1, Q^2_2)\nonumber\\
 = D^{ij}_h(x_1, x_2; Q^2_1, Q^2_2) f({\bf b_1}) f({\bf b_2}),
\end{eqnarray} 
where $f({\bf b_1})$ is supposed to be an universal function for all kind of partons with the fixed normalization,
\begin{eqnarray} 
\label{f}
\int f({\bf b_1}) f({\bf b_1 -b})d^2b_1 d^2b = \int T({\bf b})d^2b = 1,
\end{eqnarray} 
and  $T({\bf b}) = \int f({\bf b_1}) f({\bf b_1 -b})d^2b_1 $ is the overlap function. 

If one makes the further assumptions that the longitudinal components $D^{ij}_h(x_1, x_2; Q^2_1, Q^2_2)$ reduce to the product of two independent one parton distributions,
\begin{eqnarray} 
\label{DxD}
D^{ij}_h(x_1, x_2; Q^2_1, Q^2_2) = D^i_h (x_1; Q^2_1) D^j_h (x_2; Q^2_2),
\end{eqnarray}
the cross section of double parton scattering can be expressed in the simple form
\begin{eqnarray} 
\label{doubleAB}
& \sigma^D_{(A,B)} = \frac{m}{2} \frac{\sigma^S_{(A)} \sigma^S_{(B)}}{\sigma_{\rm eff}}, \\
& \sigma_{\rm eff}=[ \int d^2b (T({\bf b}))^2]^{-1}.
\end{eqnarray} 
In this representation and at the factorization of longitudinal and transverse components, the inclusive cross section of single hard scattering is written as
\begin{eqnarray} 
\label{hardS}
\sigma^S_{(A)} = \sum \limits_{i,k} \int \Gamma_{i}(x_1; {\bf b_1}; Q^2_1)
\hat{\sigma}^A_{ik}(x_1, x_1^{'}) \nonumber\\
\times\Gamma_{k}(x_1^{'}; {\bf b_1} - {\bf b}; Q^2_1) dx_1 dx_1^{'}  d^2b_1  d^2b \nonumber\\
=\sum \limits_{i,k} \int D^{i}_h(x_1; Q^2_1) f({\bf b_1})
\hat{\sigma}^A_{ik}(x_1, x_1^{'})\nonumber\\ 
\times D^{k}_{h'}(x_1^{'}; Q^2_1)f( {\bf b_1} - {\bf b}) dx_1 dx_1^{'}  d^2b_1  d^2b \nonumber\\
=\sum \limits_{i,k} \int D^{i}_h(x_1; Q^2_1)
\hat{\sigma}^A_{ik}(x_1, x_1^{'}) D^{k}_{h'}(x_1^{'}; Q^2_1) dx_1 dx_1^{'}.
\end{eqnarray}

These simplifying assumptions, even though rather common in the literature and quite convenient from a computational point of view, are not enough justified and should be revised~\cite{Blok:2010ge,Diehl:2011tt,stir}, while  the starting cross section formula~(\ref{hardAB}) was found (derived) in the momentum representation  using light-cone variables in a number of works (see, e.g., Refs.~\cite{paver,Mekhfi:1983az,Blok:2010ge,Diehl:2011tt,stir}) at the same approximations as in processes with a single hard scattering.

\maketitle
\section{\label{sec2} Momentum representation}
All the previous formulae were written in the mixed (momentum and coordinate) representation. Recall that in general for the case of the multiple parton interactions we have to use the 
{\it Generalized} Parton Distribution Functions (GPDF). In other words in the Feynman diagram (ladder) which describes the GPDF the parton momenta $k_L$ (in the left part of diagram corresponding to the amplitude $A^*$) and $k_R$ (in the right part of the diagram corresonding to amplitude $A$) may be different. Let us denote $k_L=k+q/2$ and $k_R=k-q/2$ where $q$ is the momentum transfered through the whole ladder. Since the ladders in Figs.~1 and 2 form a loop we will call $q$ --- the loop momentum. In the previous formulae instead of transverse momentum $q_t$ we had used the conjugated coordinate ${\bf b}$~\footnote{In the case of conventional DIS the cross section is given by the integral over ${\bf b}$ corresponding to $q_t=0$.}.
 
For our further goal the momentum representation is more convenient:
\begin{eqnarray}
\label{hardAB_p}
\sigma^D_{(A,B)} = \frac{m}{2} \sum \limits_{i,j,k,l} \int \Gamma_{ij}(x_1, x_2; {\bf q}; Q^2_1, Q^2_2)\nonumber\\
\times \hat{\sigma}^A_{ik}(x_1, x_1^{'}) 
\hat{\sigma}^B_{jl}(x_2, x_2^{'})\nonumber\\
\times\Gamma_{kl}(x_1^{'}, x_2^{'}; {\bf -q}; Q^2_1, Q^2_2) dx_1 dx_2 dx_1^{'} dx_2^{'} \frac{d^2q}{(2\pi)^2}.
\end{eqnarray}
The hard subprocesses {\bf A} and {\bf B} are originated by two different branches of the parton cascade. Note that only the sum of the parton momenta (in both branches) is conserved, while in each individual branch there may be some difference, $q$,
of transverse (parton) momenta in the initial state wave function and the conjugated wave function. 

\begin{figure}
\includegraphics[width=0.45\textwidth]{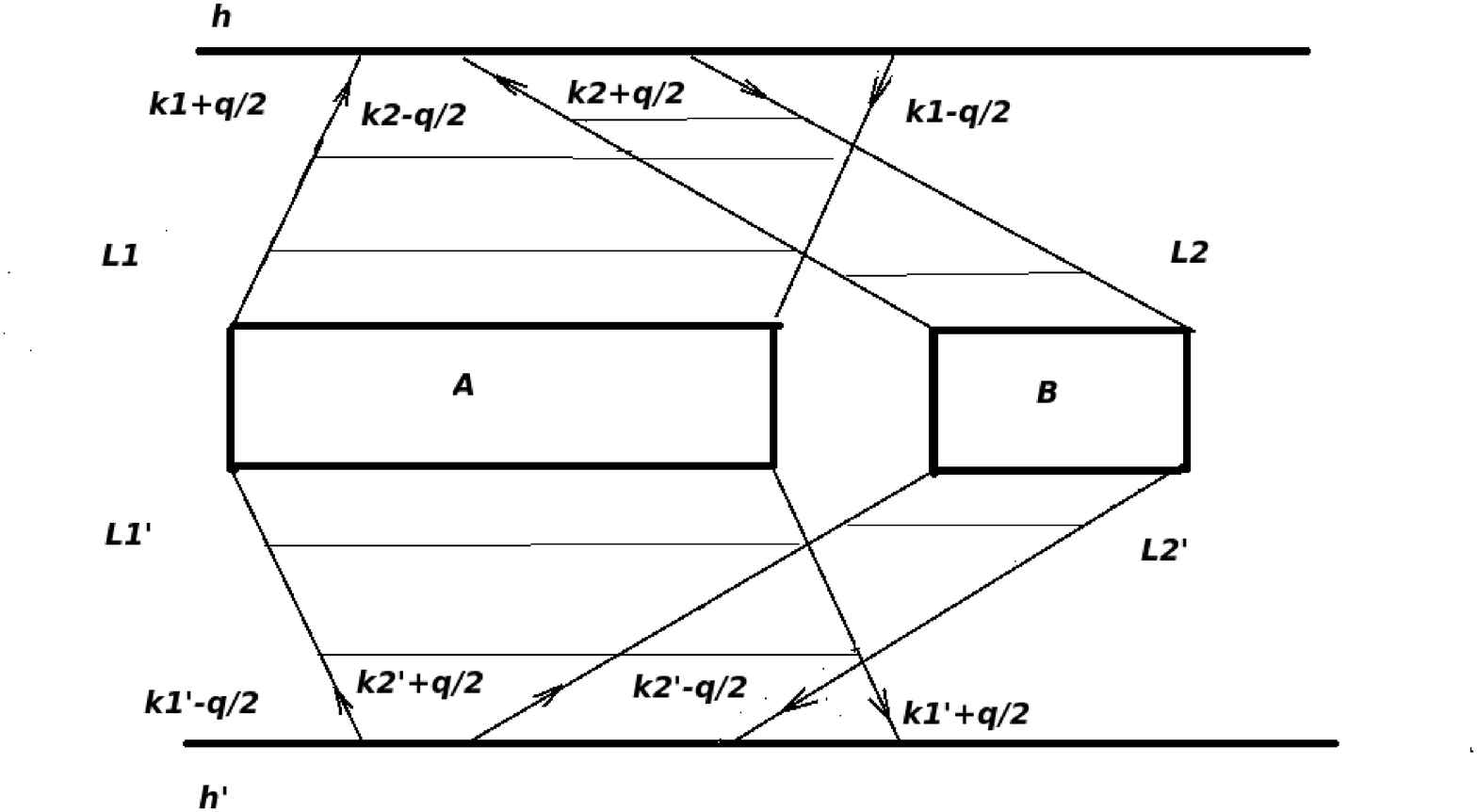}
\caption{ A graph for double parton scattering due to the first term.  $A$ and $B$ are the hard parton subprocesses. $q$ is the momemtum transfered through the ladders $L1, L2, L1'$ and $L2'$.
\label{fig:fig1}}
\end{figure}

The main problem is in the correct calculation of these two-parton functions $\Gamma_{ij}(x_1, x_2; {\bf q}; Q^2_1, Q^2_2)$ without the simplifying assumptions~(\ref{DxF}) and (\ref{DxD}). The case is that they are known in the current literature~\cite{Kirschner:1979im,Shelest:1982dg,snig03,snig04,snig08} only at ${\bf q}=0$ in the collinear approximation. In this approximation the two-parton distribution functions, $\Gamma_{ij}(x_1, x_2; {\bf q}=0; Q^2, Q^2)=D^{ij}_h(x_1, x_2; Q^2, Q^2)$ with the two hard scales set equal, satisfy the generalized Dokshitzer-Gribov-Lipatov-Altarelli-Parisi (DGLAP) evolution equations, derived for the first time in Refs.~\cite{Kirschner:1979im,Shelest:1982dg}, as well as single distributions satisfy more known and cited DGLAP equations~\cite{gribov,lipatov,dokshitzer,altarelli}. The functions in question have a specific interpretation in the leading logarithm approximation of perturbative QCD: they are inclusive probabilities that in a hadron $h$ one finds two bare partons  of  types $i$ and $j$ with the given longitudinal momentum fractions $x_1$ and $x_2$.

\begin{figure}
\includegraphics[width=0.45\textwidth]{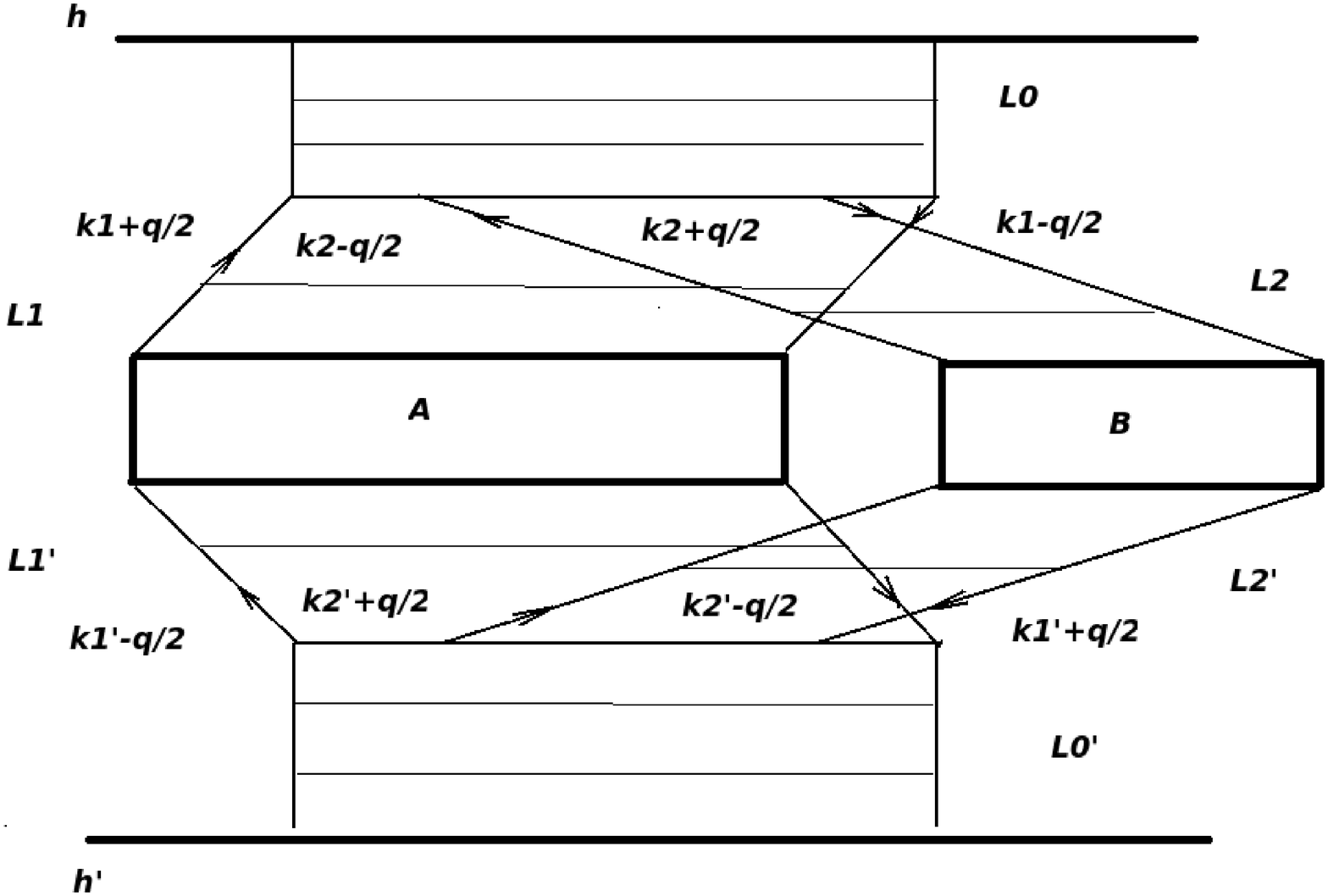}
\caption{ A graph for double parton scattering due to the second evolution term.  $A$ and $B$ are the hard parton subprocesses. $q$ is the momemtum transfered through the ladders $L1, L2, L1'$ and $L2'$
\label{fig:fig2}}
\end{figure}

The evolution equation for $\Gamma_{ij}$ contains two terms. The first term describes independent (simultaneous) evolution of two branches of parton cascade: one branch contains the parton $x_1$ and another branch --- the parton $x_2$. The second term accounts for the possibility to split the one parton evolution (one branch $k$) into the two different branches, $i$ and $j$. It contains the usual splitting function $P_{k\to ij}(z)$. The solutions of the generalized DGLAP evolution equations with the given initial conditions at the reference scales $\mu^2$ may be written~\cite{Gaunt:2009re,Ceccopieri:2010kg,snig11} 
in the form:
\begin{eqnarray}
\label{solutiontwoscale}
 D_h^{j_1j_2}(x_1,x_2;\mu^2, Q_1^2, Q_2^2)\nonumber\\
= D_{h1}^{j_1j_2}(x_1,x_2;\mu^2, Q_1^2, Q_2^2) +
D_{h2}^{j_1j_2}(x_1,x_2;\mu^2, Q_1^2, Q_2^2),\nonumber\\
 D_{h1}^{j_1j_2}(x_1,x_2;\mu^2, Q_1^2, Q_2^2)\nonumber\\ 
= \sum\limits_{j_1{'}j_2{'}} 
\int\limits_{x_1}^{1-x_2}\frac{dz_1}{z_1}
\int\limits_{x_2}^{1-z_1}\frac{dz_2}{z_2}~
D_h^{j_1{'}j_2{'}}(z_1,z_2;\mu^2)\nonumber\\
\times D_{j_1{'}}^{j_1}(\frac{x_1}{z_1};\mu^2,Q_1^2) 
D_{j_2{'}}^{j_2}(\frac{x_2}{z_2},\mu^2,Q_2^2) ,\nonumber\\
 D_{h2}^{j_1j_2}(x_1,x_2;\mu^2, Q_1^2, Q_2^2)\nonumber\\
=\sum\limits_{j{'}j_1{'}j_2{'}} \int\limits_{\mu^2}^{\min(Q_1^2,Q_2^2)}dk^2\frac{\alpha_s(k^2)}{2\pi k^2}
\int\limits_{x_1}^{1-x_2}\frac{dz_1}{z_1}
\int\limits_{x_2}^{1-z_1}\frac{dz_2}{z_2}\nonumber\\
\times D_h^{j{'}}(z_1+z_2;\mu^2,k^2) 
\frac{1}{z_1+z_2}P_{j{'} \to
j_1{'}j_2{'}}\Bigg(\frac{z_1}{z_1+z_2}\Bigg)\nonumber\\ 
\times D_{j_1{'}}^{j_1}(\frac{x_1}{z_1};k^2,Q_1^2) 
D_{j_2{'}}^{j_2}(\frac{x_2}{z_2};k^2,Q_2^2),\nonumber\\
\end{eqnarray}
where
$\alpha_s(k^2)$ is the QCD coupling, 
$D_{j_1{'}}^{j_1}(z;k^2,Q^2)$ are the known single distribution functions (the Green's functions) at the parton level with the specific $\delta$-like  initial conditions at $Q^2=k^2$. $D_h^{j'_1,j'_2}(z_1,z_2,\mu^2)$ is the initial (input) two-parton distribution at a relatively low scale $\mu$. The one parton distribution (before the slitting into the two branches at some scale $k^2$) is given by $D_h^{j'}(z_1+z_2,\mu^2,k^2)$.

Note that in Eq.~(\ref{solutiontwoscale}) we assume that the loop momentum $q<\mu$ is small and due to a strong ordering of parton  transverse momenta during the collinear DGLAP evolution it may be neglected.\\

The first term is the solution of homogeneous evolution  equation (independent evolution of two branches) where in general the input two-parton distribution at low scale $\mu$ is not known. For this non-perturbative two-parton function at low $z_1,z_2$ one may assume the factorization $D_h^{j_1{'}j_2{'}}(z_1,z_2,\mu^2) \simeq D_h^{j_1{'}}(z_1,\mu^2)D_h^{j_2{'}}(z_2,\mu^2)$
neglecting the influence of momentum conservation ($z_1+z_2<1$).

This leads to
\begin{eqnarray} 
\label{DxD_Q}
D^{ij}_{h1}(x_1, x_2; \mu^2, Q^2_1, Q^2_2)\nonumber\\
 \simeq D^i_h (x_1; \mu^2, Q^2_1) D^j_h (x_2;\mu^2, Q^2_2).
\end{eqnarray}
\noindent

Since the multiple interactions take place at relatively low transverse momenta, the contribution to the cross section from double scattering can be separated experimentally in the region of relatively low longitudinal momentum fraction $x_1$ and $x_2$, where the factorization hypothesis~(\ref{DxD_Q}) for the first term is a good approximation. In this case the cross section for double parton scattering can be estimated, using the two-gluon form factor of the nucleon $F_{2g}(q)$~\cite{Blok:2010ge,Frankfurt:2002ka} for the dominant gluon-gluon scattering mode (or something similar for other parton scattering modes),
\begin{eqnarray}
\label{hardAB1}
\sigma^{D,1\times1}_{(A,B)} = \frac{m}{2} \sum \limits_{i,j,k,l} \int D^i_h (x_1; \mu^2, Q^2_1) D^j_h (x_2;\mu^2, Q^2_2) \nonumber\\
\times \hat{\sigma}^A_{ik}(x_1, x_1^{'})
\hat{\sigma}^B_{jl}(x_2, x_2^{'}) D^k_{h'} (x'_1; \mu^2, Q^2_1)
D^l_{h'} (x'_2;\mu^2, Q^2_2)\nonumber\\
\times dx_1 dx_2 dx_1^{'} dx_2^{'} \int  F_{2g}^4(q)\frac{d^2q}{(2\pi)^2}.\nonumber\\
\end{eqnarray}

From the dipole fit $F_{2g}(q)=1/(q^2/m^2_g+1)^2$ to the two-gluon form factor follows that the characteristic value of $q$ is of the order of gluon mass $m_g$, and therefore the initial conditions for the single distributions can be fixed at some reference scale $\mu \sim m_g$ because of the weak logarithmic dependence of these distributions on the scale value. In this approach  $\int  F_{2g}^4(q)\frac{d^2q}{(2\pi)^2}$ gives the estimation of $[\sigma_{\rm eff}]^{-1}$.

The second term in Eq.~(\ref{solutiontwoscale}) is the solution of complete evolution equation with the evolution from one ``nonperturbative'' parton at the reference scale. Here the independent evolution of two branches starts at scale $k^2$ from a point-like parton $j'$. Therefore we have no the form factor $F_{2g}(q)$ which suppresses the large $q_t$ domain. The corresponding contribution to the cross section reads
\begin{eqnarray}
\label{D2xD2}
\sigma^{D,2\times2}_{(A,B)} \nonumber\\
= \frac{m}{2} \sum \limits_{i,j,k,l} \int dx_1 dx_2 dx_1^{'} dx_2^{'} \int^{\min(Q_1^2,Q_2^2)} \frac{d^2q}{(2\pi)^2}\nonumber\\
\times \sum\limits_{j{'}j_1{'}j_2{'}} \int\limits_{q^2}^{\min(Q_1^2,Q_2^2)}dk^2 \frac{\alpha_s(k^2)}{2\pi k^2}
\int\limits_{x_1}^{1-x_2}\frac{dz_1}{z_1}
\int\limits_{x_2}^{1-z_1}\frac{dz_2}{z_2}\nonumber\\
\times D_h^{j{'}}(z_1+z_2;\mu^2,k^2)
\frac{1}{z_1+z_2}P_{j{'} \to
j_1{'}j_2{'}}\Bigg(\frac{z_1}{z_1+z_2}\Bigg)\nonumber\\
\times D_{j_1{'}}^{i}(\frac{x_1}{z_1};k^2,Q_1^2) 
D_{j_2{'}}^{j}(\frac{x_2}{z_2};k^2,Q_2^2)
\nonumber\\
\times\hat{\sigma}^A_{ik}(x_1, x_1^{'}) \hat{\sigma}^B_{jl}(x_2, x_2^{'})\nonumber\\
\times\sum\limits_{j{'}j_1{'}j_2{'}} \int\limits_{q^2}^{\min(Q_1^2,Q_2^2)}dk^{'2} \frac{\alpha_s(k^{'2})}{2\pi k^{'2}}
\int\limits_{x'_1}^{1-x'_2}\frac{dz_1}{z_1}
\int\limits_{x'_2}^{1-z_1}\frac{dz_2}{z_2}\nonumber\\
\times D_{h'}^{j{'}}(z_1+z_2;\mu^2,k^{'2})
\frac{1}{z_1+z_2}P_{j{'} \to
j_1{'}j_2{'}}\Bigg(\frac{z_1}{z_1+z_2}\Bigg)\nonumber\\ 
\times D_{j_1{'}}^{k}(\frac{x'_1}{z_1};k^{'2},Q_1^2) 
D_{j_2{'}}^{l}(\frac{x'_2}{z_2};k^{'2},Q_2^2),
\end{eqnarray}
or in less transparent, but substantially shorter form:
\begin{eqnarray}
\label{D2xD2_s}
\sigma^{D,2\times2}_{(A,B)} \nonumber\\
= \frac{m}{2} \sum \limits_{i,j,k,l} \int dx_1 dx_2 dx_1^{'} dx_2^{'} \int^{\min(Q_1^2,Q_2^2)} \frac{d^2q}{(2\pi)^2}\nonumber\\
\times D_{h2}^{ij}(x_1,x_2;q^2, Q_1^2, Q_2^2)
\hat{\sigma}^A_{ik}(x_1, x_1^{'}) \nonumber\\
\times \hat{\sigma}^B_{jl}(x_2, x_2^{'})
D_{h'2}^{kl}(x'_1,x'_2;q^2, Q_1^2, Q_2^2).
\end{eqnarray}

There is yet the combined (``interference'') contribution, which is written by analogy,
\begin{eqnarray}
\label{D1xD2_s}
\sigma^{D,1\times2}_{(A,B)}\nonumber\\
= \frac{m}{2} \sum \limits_{i,j,k,l} \int dx_1 dx_2 dx_1^{'} dx_2^{'} \int^{\min(Q_1^2,Q_2^2)} F_{2g}^2(q)\frac{d^2q}{(2\pi)^2}\nonumber\\
\times [D^i_h (x_1; \mu^2, Q^2_1) D^j_h (x_2;\mu^2, Q^2_2)
\hat{\sigma}^A_{ik}(x_1, x_1^{'}) \nonumber\\
\times \hat{\sigma}^B_{jl}(x_2, x_2^{'})
D_{h'2}^{kl}(x'_1,x'_2;q^2, Q_1^2, Q_2^2)\nonumber\\
+D_{h2}^{ij}(x_1,x_2;q^2, Q_1^2, Q_2^2)
\hat{\sigma}^A_{ik}(x_1, x_1^{'}) \nonumber\\
\times \hat{\sigma}^B_{jl}(x_2, x_2^{'})
D^k_{h'} (x'_1; \mu^2, Q^2_1) D^l_{h'} (x'_2;\mu^2, Q^2_2)].\nonumber\\
\end{eqnarray}

The equations~(\ref{hardAB1}), (\ref{D2xD2_s}) and  (\ref{D1xD2_s}) are our solution of the problem --- the estimation of the inclusive cross section for double parton scattering, taking into account the QCD evolution, in terms of the well-known collinear distributions extracted from deep inelastic scattering.
However, here one should note that  
the input two-parton distribution $D_h^{j'_1,j'_2}(z_1,z_2,\mu^2)$ may be more complicated than that given by factorization ansatz (\ref{DxD_Q}). Let us discuss in more detail the second term, that is the $2\times 2$ contribution.

\maketitle
\section{\label{sec3} Discussion and conclusions}
The contribution to the cross section from the second term induced by the QCD evolution does not reduce to the simple contribution with a some new, other, constant effective cross section $\sigma_{\rm eff}$, as it was done in first estimations~\cite{Cattaruzza:2005nu,Gaunt:2010pi,Maina:2010vh}. The QCD evolution effects for  the cross section are anticipated to be larger than for the two-parton distribution functions, for which they were estimated in Refs.~\cite{Gaunt:2009re,snig04} on the level of 10$\%$ -30$\%$ in comparison with the ``factorization'' components at $x \sim 0.1$ and $Q \sim 100$~GeV. Indeed, in Eq.~(\ref{D2xD2}) the integration over $q$ has no a strong suppression factor $F_{2g}(q)$ and the  phase space integral may be estimated as,
\begin{eqnarray}
\label{pspace}
\int^{Q^2} dq^2 \int_{q^2}^{Q^2} \frac{dk^2}{k^2}\int_{q^2}^{Q^2} \frac{dk^{'2}}{k^{'2}} \simeq 2Q^2 ,~~~Q^2 \gg \mu^2, 
\end{eqnarray}
where within the leading order (LO) accuracy we take $q^2$ as the lower limit in $k^2$ and $k^{'2}$ integrations; at $q^2>k^2$ the loop momentum $q_t$ destroys the logarithmic structure of the integrals in collinear evolution from $k^2$ to $Q^2$.

We see that for a large final scale $Q^2$ the second ($2\times 2$) contribution dominates being proportional to $Q^2$ in comparison with the $1\times 1$ or $1\times 2$ components which contributions $\sim m^2_g\sim 1/\sigma_{eff}$ are limited by the nucleon (hadron) form factor $F_{2g}$~\footnote{In terms of impact parameters ${\bf b}$ this means that in the second ($2\times 2$) term two pairs of partons are very close to each other; $|{\bf b}_1-{\bf b}_2|\sim 1/Q$.}.

The real gain is, of course, less, since the running coupling constant and the distribution functions are logarithmically dependent on the integration variables and we have the additional suppression factor, which is inversely proportional~\cite{Snigirev:2010ds} to the initial gluon and quark multiplicities squared; due to the different ``normalization'': the second term evolves from  one ``nonperturbative''  parton unlike the first factorization term having the two initial independent ``nonperturbative''  partons at the reference scale.

As a result, the experimental effective cross section, $\sigma_{\rm eff}^{\rm exp}$, which is 
not measured directly but is extracted, using the normalization to the product of two single cross sections:
\begin{eqnarray} 
\label{dps}
\frac{\sigma_{DPS}^{\gamma+3j}}{\sigma^{\gamma j}\sigma^{jj}}= [\sigma_{\rm eff}^{\rm exp}]^{-1}
\end{eqnarray} 
in both the CDF and D0 experiments, should be dependent on the probing hard scale. It should decrease with the growth of the resolution scale due to the fact that all additional contributions to the cross section of double parton scattering are positive and increase. Here $\sigma^{\gamma j}$ and $\sigma^{j j}$ are the inclusive $\gamma +$ jet and dijets cross sections, $\sigma_{DPS}^{\gamma+3j}$ is the inclusive cross section of the $\gamma + 3$ jets events produced in the double parton process. It is worth noticing that the CDF and D0 Collaborations extract $\sigma_{\rm eff}^{\rm exp}$ without any theoretical predictions on the $\gamma +$ jet and dijets cross sections, by comparing the number of observed double parton $\gamma + 3$ jets events in one $p{\bar p}$ collision to the number of $\gamma + 3$ jets events with hard interactions occurring in the case of two separate  $p{\bar p}$ collisions.

The recent D0 measurement~\cite{D0} of this effective cross section, $\sigma_{\rm eff}^{\rm exp}$, done as a function of the second (ordered in the transverse momentum $p_T$) jet $p_T$, $p^{\rm jet2}_T$, that can serve as a resolution scale, shows a tendency to be dependent on this scale. In Ref.~\cite{Snigirev:2010tk}
this fact was interpreted as the first indication to the QCD evolution of double parton distributions.\\

We have to emphasize that the dominant contribution to the phase space integral~(\ref{pspace}) comes from a large $q^2\sim Q^2$ and strictly speaking we have no place for the collinear (DGLAP) evolution of two independent branches of the parton cascade (i.e., in the ladders $L1, L2, L1'$ and $L2'$) in the $2\times 2$ term. Formally in the framework of collinear approach this contribution should be considered as that caused by the interaction of {\it one} pair of partons with the $2\to 4$ hard subprocess~\footnote{This is in agreement with the statement~\cite{stir} that ``the structure of Fig.~2 should not be included in the leading logarithmic {\it double}  parton scattering cross section''.}. Recall however that in the estimate (\ref{pspace}) we neglect the anomalous dimension, $\gamma$, of the parton distributions $D^k_j(x/z,k^2,Q^2)\propto (Q^2/k^2)^\gamma$. In collinear appoach the anomalous dimensions
$\gamma\propto \alpha_s<<1$ are assumed to be small. On the other hand  in a low $x$ region the value of anomalous dimension is enhanced by the $\ln(1/x)$ logarithm and may be rather large numerically. So the integral over $q^2$ is {\it slowely} convergent and the major contribution to the cross section is expected to come actually from some characteristic intermediate region, $m^2_g<<q^2<< Q^2_1 (Q_1<Q_2)$, with the not such strong sensitivity to the upper limit of $q$-integration as in the case of the phase space integral. Therefore it makes sense to consider the numerical contribution of the $2\times 2$ term even within the collinear approach at the LHC kinematics with the available large enough $Q_1$ and $Q_2$ with respect to $m_g$ (providing the wide enough integration region and the characteristic intermediate momenta $q$).

Next in a configuration with two quite different scales (say, $Q^2_1<<Q^2_2$) the upper limit of $q^2$ integral is given by a smaller scale (at $q>Q_1$ the hard matrix element corresponding to $\sigma^A$ starts to fall down with $q_t$). In this case the collinear evolution from the scale $q=Q_1$ up to the scale $Q_2$  in the ladders (parton branches) $L2$ and $L2'$ is well justified.\\

Moreover the integrals over $z_1$, $z_2$ in Eq.~(\ref{D2xD2}) are concentrated at a rather large $z$ giving enough space for the Balitsky-Fadin-Kuraev-Lipatov (BFKL) evolution~\cite{bfkl,bfkl2,bal,ryskin}. Thus it may be intersting (and even more justified theoretically) to consider the multiple parton interactions in the framework of the BFKL approach (see the recent paper~\cite{flesburg}).\\

In summary, we suggest a practical constructive me\-thod how to estimate the inclusive cross section for double parton scattering, taking into account the QCD evolution, in terms of the well-known collinear distributions extracted from deep inelastic scattering. We support also the conclusion of Refs.~\cite{Snigirev:2010tk,flesburg} that the experimentally measured effective cross section, $\sigma_{\rm eff}^{\rm exp}$, at a normalization~(\ref{dps}) and the presence of the evolution (correlation) term in the two-parton distributions, should decrease with the growth of the resolution scale $Q^2$.

\begin{acknowledgments}
Discussions with D.V.~Bandurin, E.E.~Boos, S.V.~Mo\-lo\-dtsov, M.~Strikman, D.~Treleani, G.M.~Zinovjev and N.P.~Zotov are gratefully acknowledged. This work is partly supported by Russian Foundation for Basic Research Grants  No. 10-02-93118, the President of Russian Federation for support of Leading Scientific Schools Grant No 4142.2010.2. and by the Federal Program of the Russian State
RSGSS=65751.2010.2.
\end{acknowledgments}


\end{document}